\documentclass[aps,prd,nofootinbib,twocolumn,superscriptaddress,floatfix,notitlepage]{revtex4-1}
\pdfoutput=1

\usepackage[normalem]{ulem}
\usepackage{graphicx}
\usepackage{amsmath,amsfonts,amssymb}
\usepackage[dvipsnames]{xcolor}
\usepackage[breaklinks,colorlinks,urlcolor=blue,citecolor=blue,linkcolor=magenta]{hyperref}
\usepackage{enumitem}
\usepackage{slashed}

\newcommand{\be}{\begin{equation}} 
\newcommand{\ee}{\end{equation}}
\newcommand{\bea}{\begin{equation}\begin{aligned}} 
\newcommand{\eea}{\end{aligned}\end{equation}}
\newcommand{\Hubble}{\mathcal{H}}
\newcommand{\td}{{\rm d}}

\begin{document}

\title{Irreducible cosmological backgrounds of a real scalar with a broken symmetry}

\author{Francesco D'Eramo}
\email{francesco.deramo@pd.infn.it}
\affiliation{Dipartimento di Fisica e Astronomia, Universit\`a degli Studi di Padova, Via Marzolo 8, 35131 Padova, Italy}
\affiliation{Istituto Nazionale di Fisica Nucleare, Sezione di Padova, Via Marzolo 8, 35131 Padova, Italy}

\author{Andrea Tesi}
\email{andrea.tesi@fi.infn.it}
\affiliation{Dipartimento di Fisica e Astronomia, Universit\`a degli Studi di Firenze, Via Sansone 1, 50019 Sesto Fiorentino, Italy}
\affiliation{Istituto Nazionale di Fisica Nucleare, Sezione di Firenze, Via Sansone 1, 50019 Sesto Fiorentino, Italy}

\author{Ville Vaskonen}
\email{ville.vaskonen@pd.infn.it}
\affiliation{Dipartimento di Fisica e Astronomia, Universit\`a degli Studi di Padova, Via Marzolo 8, 35131 Padova, Italy}
\affiliation{Istituto Nazionale di Fisica Nucleare, Sezione di Padova, Via Marzolo 8, 35131 Padova, Italy}
\affiliation{Keemilise ja bioloogilise f\"u\"usika instituut, R\"avala pst. 10, 10143 Tallinn, Estonia}

\begin{abstract}
We explore the irreducible cosmological implications of a singlet real scalar field. Our focus is on theories with an approximate and spontaneously broken $\mathbb{Z}_2$ symmetry where quasi-stable domain walls can form at early times. This seemingly simple framework bears a wealth of phenomenological implications that can be tackled by means of different cosmological and astrophysical probes. We elucidate the connection between domain wall dynamics and the production of dark matter and gravitational waves. In particular, we identify three main benchmark scenarios. The gravitational wave signal observed by pulsar timing arrays can be generated by the domain walls if the mass of the singlet is $m_s \sim\,$PeV. For lower masses, but with $m_s \gtrsim 10\,$GeV, scalars produced in the annihilation of the domain walls can be dark matter with a distinctive feature in their power spectrum. Finally, the thermal bath provides an unavoidable source of unstable scalars via the freeze-in mechanism whose subsequent decays can be tested by their imprints on cosmological and terrestrial observables.
\end{abstract}

\maketitle

\section{Introduction}
\label{sec:intro}

The observed evidence for dark matter (DM) at vastly different astrophysical and cosmological length scales provides undeniable evidence for the need for physics beyond the Standard Model (SM)~\cite{Bertone:2004pz,Cirelli:2024ssz}. The lack of new physics signals at the Large Hadron Collider (LHC) and at direct and indirect searches has been putting weak scale models under siege in the last few years~\cite{Arcadi:2024ukq}. New feebly coupled degrees of freedom arise naturally in motivated frameworks for physics beyond the SM, and in some regions of the parameter space are viable DM candidates. They are very difficult to look for in terrestrial experiments due to their tiny interactions with visible particles. However, the early Universe provides a unique laboratory to investigate this kind of theoretical framework because of the high energies and/or densities achieved across the expansion history.

In this work, we reconsider the simplest (in terms of the number of new degrees of freedom) SM extension: the addition of a singlet real scalar field $S$. Within this framework, which has been studied in various contexts, the only renormalizable interactions for the new field are with the SM Higgs doublet $H$ and appear in the scalar potential. The case where the theory has a $\mathbb{Z}_2$ symmetry, with $S$ the only field that is odd under it, has all the ingredients to provide a viable DM candidate if $S$ not develop a vacuum expectation value (vev)~\cite{McDonald:1993ex,Burgess:2000yq,McDonald:2001vt,Barger:2008jx,Djouadi:2011aa,Cline:2012hg,Cline:2013gha,Gabrielli:2013hma,GAMBIT:2017gge,Bernal:2018kcw,Fairbairn:2018bsw,Kannike:2019mzk,Chiang:2019oms,Tenkanen:2019aij,Alanne:2020jwx,Coito:2021fgo}. Another interesting case is for Higgs portal interactions large enough and scalar masses near the electroweak scale since this new degrees of freedom can enable a first-order electroweak phase transition needed for electroweak baryogenesis~\cite{McDonald:1993ey,Espinosa:1993bs,Choi:1993cv,Ham:2004cf,Espinosa:2007qk,Profumo:2007wc,Ahriche:2007jp,Ahriche:2012ei,Alanne:2014bra,Profumo:2014opa,Alanne:2016wtx,Tenkanen:2016idg,Vaskonen:2016yiu,Beniwal:2017eik,Ellis:2022lft}. Despite its simplicity, the real scalar appears in countless low energy spectra of more elaborated models and it offers a proxy to study topological defects that can form in the early Universe \cite{Zeldovich:1974uw,Kibble:1976sj,Vilenkin:1984ib,Gelmini:1988sf,Coulson:1995nv,Vilenkin:2000jqa}.

Our focus is on the scenarios where this framework has an approximate discrete $\mathbb{Z}_2$ symmetry. We consider the parameter space region where, even in the absence of the tiny $\mathbb{Z}_2$-breaking terms, the discrete symmetry is not respected by the global minimum of the zero temperature scalar potential. We describe how this spontaneous breaking is responsible for the formation of cosmological defects in the form of \textit{domain walls} (DW). The DW network is long-lived and acquires a large energy density that it will eventually release when explicit $\mathbb{Z}_2$-breaking terms become important. The production of gravitational waves (GWs) and other light quanta at such a time is believed to be sizeable~\cite{Vilenkin:1981zs,Lyth:1991bb,Nagasawa:1994qu,Gleiser:1998na,Hiramatsu:2010yn,Hiramatsu:2010yz,Hiramatsu:2012gg,Hiramatsu:2012sc,Hiramatsu:2013qaa,Kawasaki:2014sqa,Chang:2023rll}. The spontaneous breaking is the main source that renders the scalar unstable, and depending upon the size of the portal couplings it can decays during the first stages of the expansion history or even be cosmologically stable. Both these limits are interesting: the former allows us to constrain the scenario if the scalar was produced in the early universe while the latter opens up the possibility for the scalar to contribute to the observed DM abundance.

We introduce the theoretical framework in Sec.~\ref{sec:theory}. In particular, we identify the $\mathbb{Z}_2$-preserving and violating parts in the scalar potential. The latter is a sub-dominant correction that becomes relevant only when we consider the annihilation of DW. Thus we discuss the mass spectrum, the scalar lifetime, and the thermal corrections in the $\mathbb{Z}_2$-preserving limit. The dynamics of DW production, both thermal and non-thermal, and annihilation are discussed in Sec.~\ref{sec:DW}. The parameter space is rather broad, and depending on the size of the scalar mass and the portal coupling with the SM Higgs doublet we have rather distinct phenomenology that spans from GW production to DM production. In the intermediate region where the new scalar is not enough long-lived to account for DM and the GW signal is not detectable, there is an unavoidable production from the thermal bath via the freeze-in mechanism that can leave an imprint on cosmological and terrestrial observables. Thus we identify the following three main benchmark scenarios.
\begin{itemize}
    \item \textbf{Gravitational Waves.} If DWs annihilate quickly we can target the parameter space with GWs. Several pulsar timing array (PTA) collaborations have recently reported evidence for a GW background at nHz frequencies~\cite{NANOGrav:2023gor,EPTA:2023fyk,Reardon:2023gzh,Xu:2023wog} and various groups have studied if that signal could arise from DW annihilation~\cite{Ferreira:2022zzo,NANOGrav:2023hvm,Ellis:2023oxs,Gouttenoire:2023ftk,Babichev:2023pbf,Gelmini:2023kvo}. As shown in Ref.~\cite{Ellis:2023oxs}, the stochastic GW background generated by a DW network gives, in fact, a relatively good fit to the NANOGrav 15-year data. This is the subject of Sec.~\ref{sec:GWs}.
    \item \textbf{Dark Matter.} DW annihilation can produce a population of cosmologically stable scalars. In this case, DM may be produced at rather late times, with temperatures as low as the one of matter-radiation equality, and we show how matter power spectrum data can strongly constrain this scenario. This analysis can be found in Sec.~\ref{sec:DM}. 
    \item \textbf{Unavoidable Freeze-In.} If DW annihilation is not phenomenologically relevant, we explore the impact of the new scalar in a more standard scenario where its interactions allow for freeze-in production via decays and scatterings involving thermal bath particles. Depending on the size of the portal coupling, we explore different bounds from Big Bang Nucleosynthesis (BBN), the Cosmic Microwave Background (CMB), and X-ray searches. This unavoidable source of scalars and the associated constraints are described in Sec.~\ref{sec:thermal}.
\end{itemize}
The content of the three sections described above is rather technical. Readers not interested in technicalities can safely jump to the summary provided in the final Sec.~\ref{sec:conclusions} where the main results of this work are compactly summarized by the plot in Fig.~\ref{fig:summary}.

\section{The Framework}
\label{sec:theory}

In this section, we introduce the particle physics framework that is the subject of our study. The only new degree of freedom that we add to the SM field content is a real scalar $S$, and we assume it to be a singlet under the SM gauge group. Within this simple framework, the full Lagrangian takes the schematic form
\be
    \mathcal{L} = \mathcal{L}_{\rm SM} + \frac{1}{2} \left(\partial_\mu S\right)^2 + V_{\mathbb{Z}_2}(H,S) + V_{\not{\mathbb{Z}_2}}(H,S) + V_0 \,.
\ee
The first term on the right-hand side is the well-known SM Lagrangian, and it contains the pure SM potential terms for the Higgs doublet field $H$ 
\be
   - \mathcal{L}_{\rm SM} \supset V_{\rm SM}(H) = -\mu_H^2 |H|^2 + \lambda_H |H|^4 \,.
\ee
This part of the SM Lagrangian is particularly relevant to the upcoming analysis because the gauge neutrality of $S$ allows for several new interactions but only in the scalar potential. We introduce a $\mathbb{Z}_2$ symmetry to classify them, and we impose that all SM fields are even under it while $S$ is odd. The $\mathbb{Z}_2$-preserving contribution reads
\be
    V_{\mathbb{Z}_2}(H,S) = - \frac{\mu_S^2}{2} S^2 + \frac{\lambda_S}{4} S^4 + \frac{\lambda_P}{2} S^2 |H|^2 \,.
\label{eq:VZ2}
\ee
Up to a tadpole term for $S$, that can be redefined away, we have also a $\mathbb{Z}_2$-breaking part
\be
    V_{\not{\mathbb{Z}_2}}(H,s) =  \mu_1 \, S |H|^2 - \frac{\mu_3}{3} S^3 \,.
\ee
Finally, the constant $V_0$ is chosen to make the potential vanishing once evaluated in the global minimum. The requirement that the potential is bounded from below translates into $\lambda_H > 0$, $\lambda_S > 0$, and $\lambda_P > -2 \sqrt{\lambda_H \lambda_S}$.

\begin{figure}
\centering
\includegraphics[width=0.86\columnwidth]{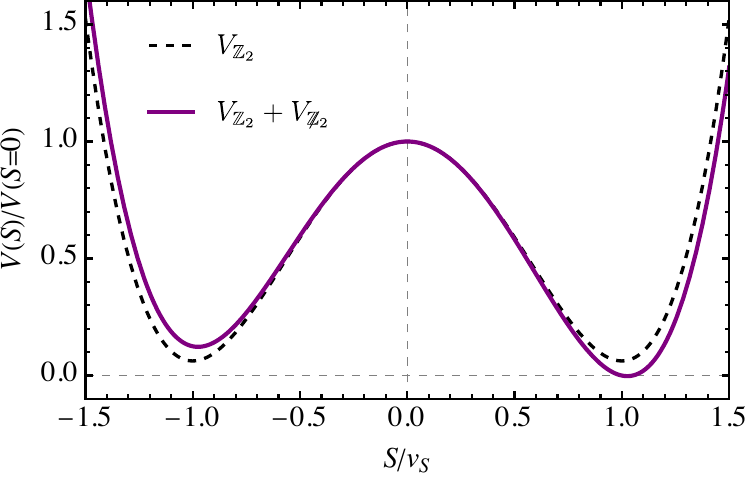}
\caption{An illustration of the scalar potential for $S$ in the limit when the portal quartic interaction with the SM Higgs doublet is suppressed, $\lambda_P \ll 1$. The dashed line shows the $\mathbb{Z}_2$-symmetric contribution whereas the solid line includes the bias term $\mu_3 = 0.05 \, \lambda_S v_S$ (we set $\mu_1 = 0$). The constant $V_0$ is set to have a vanishing potential when the scalar field sits at the positive minimum, $V(v_S > 0) = 0$.}
\label{fig:VR}
\end{figure}

The focus of this work is on the parameter space region where the scalar potential supports the formation of quasi-stable DWs. This requires two conditions: First, the field $S$ needs to develop a non-vanishing vev even in the $\mathbb{Z}_2$-symmetric limit. Thus, the discrete symmetry needs to be spontaneously broken. Before we add $\mathbb{Z}_2$-breaking terms, the scalar potential has two degenerate minima for $\langle S \rangle = \pm v_S$. If we do not consider other potential terms, the DWs are cosmologically stable and the scenario is not viable. Thus, the second condition is an explicit breaking of the $\mathbb{Z}_2$ symmetry via the $\mu_1$ or $\mu_3$ bias term that lifts the degeneracy between the minima. For simplicity, we consider only a non-zero $\mu_3$ and set $\mu_1 = 0$. The effect of this bias on the scalar potential part that depends on $S$ only for $\mu_3 = 0.05 \, \lambda_S v_S$ and $\lambda_P \ll 1$ is illustrated in Fig.~\ref{fig:VR}. 

As a side note, we observe that DWs can form also for parameter space regions where $S$ has a vanishing vev as long as $\lambda_P < -3\lambda_S/2$ (see Eq.~\eqref{eq:thermalcorrections}). However, they exist only before the electroweak phase transition and can trigger it~\cite{Blasi:2022woz,Agrawal:2023cgp,Wei:2024qpy}. Here, we focus instead on the case $\langle S \rangle \neq 0$ where the DWs can be long-lived. 

In the remainder of this section, we will discuss the scalar mass spectrum with a focus on the mixing between the radial model of the SM Higgs doublet $H$ and $S$. The resulting mixing angle $\theta$ controls the lifetime of the mostly-singlet scalar mass eigenstate. We will list all possible decays modes, evaluate the branching ratios, and quantify the lifetime. Finally, we will discuss briefly the thermal corrections to the scalar potential that are relevant for the continuation of this analysis. We will always work in the limit where the bias term $\mu_3$ is much smaller than the other dimensionful parameters in the scalar potential. In other words, we will discuss spectrum and interactions in the $\mathbb{Z}_2$-symmetric limit.

\subsection{Mass spectrum}

We expand both scalars around their vevs, and we take the unitary gauge for the SM Higgs doublet
\bea
    H = \frac{1}{\sqrt{2}} \left( \begin{array}{c} 0 \\ v_H + h^\prime \end{array} \right) \ , \qquad \qquad 
    S = v_S + s^\prime \ .
\eea
The primed notation is to save the symbols without the prime for mass eigenstates. Keeping only $\mathbb{Z}_2$-preserving terms, the explicit expressions for the vevs read
\bea \label{eq:vevs}
    v_H = & \, \frac{\mu_H}{\sqrt{\lambda_H}} \left( 1 - \frac{\lambda_P}{2 \lambda_S} \frac{\mu_S^2}{\mu_H^2} \right)^{1/2} \left(1 - \frac{\lambda_P^2}{4\lambda_H \lambda_S} \right)^{-1/2} \ ,  \\ 
    v_S = & \, \frac{\mu_S}{\sqrt{\lambda_S}} \left( 1 - \frac{\lambda_P}{2 \lambda_H} \frac{\mu_H^2}{\mu_S^2} \right)^{1/2} \left(1 - \frac{\lambda_P^2}{4\lambda_H \lambda_S} \right)^{-1/2} \ .
\eea
The limiting expressions valid for small portal couplings are easily obtained from the results above.

The mass matrix for the scalar fields contains a mixing term between $h^\prime$ and $s^\prime$ that is proportional to the quartic coupling, and it explicitly reads
\be
    M^2 = \left( \begin{array}{cc}
    2 \lambda_H v_H^2 & \lambda_P v_H v_S \\ 
    \lambda_P v_H v_S & 2 \lambda_S v_S^2
    \end{array}
    \right) \ .
\ee
The off-diagonal terms go to zero if we switch off the portal interaction. We introduce the mass eigenstates
\be
    \left(\begin{array}{c} 
    h^\prime \\ s^\prime
    \end{array}\right) = \left( \begin{array}{cc}
    \cos\theta & - \sin\theta \\ 
    \sin\theta & \cos\theta 
    \end{array}
    \right) \left(\begin{array}{c} 
    h \\ s
    \end{array}\right)
\ee
For a two-dimensional mass matrix the mixing angle $\theta$ can be found analytically and it results in
\be \label{eq:theta}
    \tan(2\theta) = \frac{\lambda_P v_H v_S}{\lambda_H v_H^2 - \lambda_S v_S^2} \ .
\ee

A sizable mixing angle affects Higgs couplings and impacts collider phenomenology (see, e.g., Refs.~\cite{Robens:2015gla,Robens:2016xkb}). In this work, we always work in the $\theta \ll 1$ so that the modifications to the couplings of the SM-like Higgs boson $h$ are harmless. A small mixing angle can be achieved either because the portal coupling $\lambda_P$ is small or there is a large hierarchy between $v_S$ and $v_H$. Either way, the mass spectrum in this regime results in 
\bea \label{eq:spectrum}
    m_h \approx & \, \sqrt{2\lambda_H} \, v_H \ , \\
    m_s \approx & \, \sqrt{2\lambda_S} \, v_S \ .
\eea
With the only exception of the region $m_h \simeq m_s$ where $\theta = \pi/4$, in the small mixing angle regime we can use the approximate expression
\be \label{eq:angle}
    \theta \approx \frac{\lambda_P}{2 \sqrt{\lambda_H \lambda_S}}  \frac{m_h m_s}{m_h^2 - m_s^2} \ .
\ee

\subsection{Singlet decay modes and lifetime}

The mostly-singlet mass eigenstate $s $ is unstable even in the $\mathbb{Z}_2$-symmetric limit as a consequence of the non-vanishing $v_S$. We evaluate its lifetime and quantify the branching ratios for its decay modes. If one looks at the scalar potential, the only possible final state is a pair of Higgs bosons (kinematically allowed if $m_s > 2 m_h$). The operator mediating this decay reads
\be
    V_{shh} = D_{shh} \, \lambda_P v_S \, s \frac{h^2}{2}
\ee
and the explicit expression for the dimensionless coefficient $D_{shh}$ is obtained by expressing the scalar potential in the mass eigenbasis
\bea
    D_{shh} = \cos^3\theta & \, \left[1 + 2 \frac{v_H}{v_S} \left(1 - 3 \frac{\lambda_H}{\lambda_P}\right) \tan\theta + \right. \\ & \left. - 2 \left(1 - 3 \frac{\lambda_S}{\lambda_P} \right) \tan^2\theta - \frac{v_H}{v_S} \tan^3\theta \right] \ .
\eea
The corresponding decay width reads
\be \label{eq:Gammashh}
    \Gamma_{s \rightarrow h h} = D_{shh}^2 \, \frac{\lambda_P^2 v_S^2}{32 \pi m_s} \sqrt{1-\frac{4m_h^2}{m_s^2}} \ .
\ee
The field $s$ decays to other SM particles through the mixing. For each final state $XX$ of SM Higgs boson decays, the partial width for $s\to XX$ is given by
\be
    \Gamma_{s\to XX} = \sin^2\theta \, \Gamma_{h\to XX}(m_s) 
\ee
with $\Gamma_{h\to XX}(m_s)$ the SM Higgs boson partial decay width for the $XX$ final state as a function of its mass~\cite{Gunion:1989we,Djouadi:2005gi}. We employ the estimates derived in~\cite{Winkler:2018qyg} for the decays to pions and kaons and include the QCD corrections derived in~\cite{Drees:1990dq} for quark final states. 

\begin{figure*}
\centering
\includegraphics[height=0.37\textwidth]{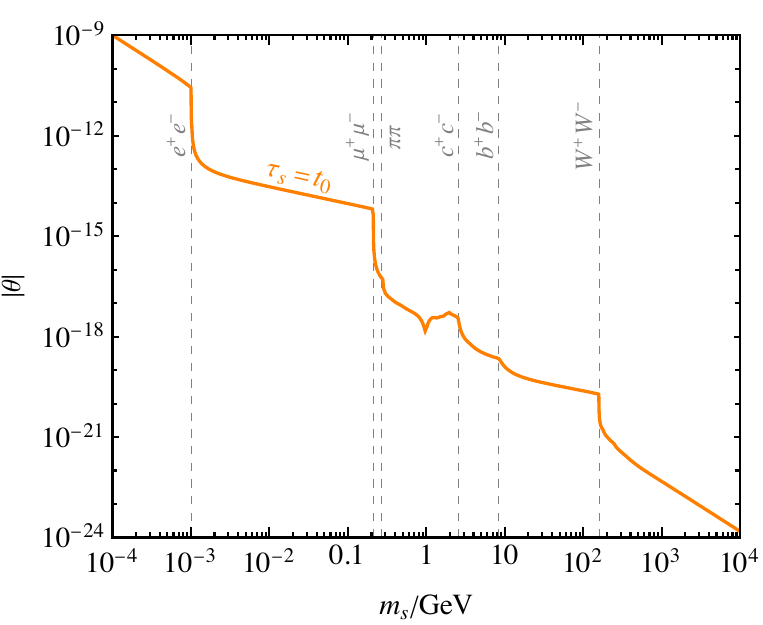} \hspace{2mm}
\includegraphics[height=0.37\textwidth]{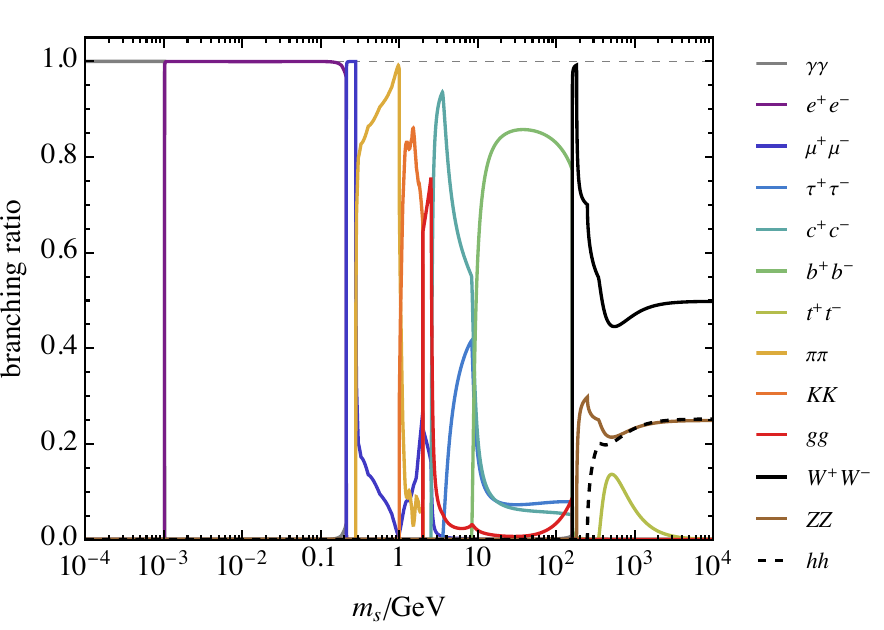}
\caption{\emph{Left panel:} Absolute value of the mixing angle $\theta$ as a function of $m_s$ needed to ensure that the mostly-singlet $s$ lifetime equals the present age of the Universe. The vertical dashed lines identifies the masses where some decay channels open up. \emph{Right panel:} Branching ratios of $s$ in the small mixing angle limit.}
\label{fig:theta}
\end{figure*}

To summarize, the total decay width reads
\be
    \Gamma_s = \Gamma_{s \rightarrow h h} + \sin^2\theta \, \Gamma_{h}(m_s) \,.
    \label{eq:Gammas}
\ee
The lifetime of $s$, which is defined as $\tau_s = 1/\Gamma_s$, depends only on the absolute value of the mixing angle $\theta$ and not on its sign. The orange line in the left panel of Fig.~\ref{fig:theta} shows for what values of $m_s$ and $|\theta|$ the lifetime of $s$ equals the present age of the Universe, $t_0 = 13.8\times 10^9$\,yr. For the parameter space region above the orange curve, $s$ is not a viable DM candidate because it would not be present anymore today. In the region below, the lifetime of $s$ is long enough for it to be present until today. However, as we will discuss in Sec.~\ref{sec:DM}, it is not enough to impose that $s$ survives until today because there are further constraints due to late-time decays of $s$ affecting the CMB anisotropy spectrum and leading to signals in current searches. The discussion of these bounds require the knowledge of the dominant decay channel which dependents on what mass region we are investigating. To ease this analysis, we show in the left panel of Fig.~\ref{fig:theta} the branching ratios as a function of $m_s$.

We conclude this discussion with analytical estimates valid in the high mass limit, $m_s \gg m_h$, where $\Gamma_h(m_s)$ in Eq.~\eqref{eq:Gammas} is dominated by decays to weak gauge bosons. For this specific channel we find the expression
\be
    \Gamma_{s \rightarrow V V} \simeq \theta^2 \, \frac{G_F m_H^3}{16 \sqrt{2} \pi} \, \delta_V \simeq 
    \frac{\lambda_P^2 v_S^2}{32 \pi m_s} \delta_V \ ,
\ee
where $\delta_V = \{1, 2\}$ for $V = \{Z, W\}$. We notice that we recover the relations $\Gamma_{s \rightarrow W^+ W^-} = 2 \, \Gamma_{s \rightarrow Z Z} = 2 \, \Gamma_{s \rightarrow h h}$ for such high masses. These results are consistent with scalar decays mediated by the portal operator proportional to $\lambda_P$, and the final state weak bosons identified as the would-be Goldstone bosons that provide the longitudinal polarizations to the weak gauge bosons.

\subsection{Thermal corrections}

If the scalar field $S$ reaches equilibrium with the SM bath, it gets thermal corrections and it also affects the ones for the Higgs field. At leading order, these can be approximated with thermal mass terms
\bea \label{eq:muT}
    \mu_S(T)^2 = & \, \mu_S^2 - c_S T^2 \ , \\
    \mu_H(T)^2 = & \, \mu_H^2 - c_H T^2 \ ,
\eea
where
\bea
    c_S &= \frac{1}{12}(2\lambda_P + 3 \lambda_S) \ , \\
    c_H &= \frac{1}{48}(9 g_L^2+3g_Y^2+12 y_t^2+24\lambda_H+ 2\lambda_P) \ .
    \label{eq:thermalcorrections}
\eea
These corrections stabilize the Higgs field to $|H|=0$ at high temperatures. Furthermore, the singlet field is also stabilized to $|S|=0$ at high $T$ if $\lambda_P > -3\lambda_S/2$. As already mentioned, this is the parameter space region investigated in this work.

\section{Domain walls dynamics}
\label{sec:DW}

DWs are field configurations where $S$ interpolates between the two minima located at $S \approx \pm v_S$, and they are realized if $S$ ends up in different minima of its potential in different patches of the Universe. This section is devoted to discussing the dynamics of these field configurations. In particular, we will explain how DWs can be formed in the early Universe, and how the bias term provides a mechanism for their annihilation.

\subsection{DW Production I: Thermal}

If the field $S$ reaches thermal equilibrium with the SM bath then the DW production mechanism is thermal. We notice how thermal production happens even if $S$ is decoupled from the SM bath but is able to self-thermalize with a different temperature. One example of this latter case is for inflaton decays into relativistic $S$ particles and the self-coupling of $S$ is sufficiently strong. 

In this study, we consider only the former scenario where $S$ thermalizes with the SM bath. Above the weak scale, the interaction rate reads approximately $\Gamma_P = n_S \langle \sigma v \rangle \simeq (\zeta(3) \lambda_P^2 T) / (16 \pi^3)$. This has to be compared with the expansion rate quantified by the Hubble parameter $\Hubble \simeq [\pi \sqrt{g_*(T)}/ (3 \sqrt{10})] \, T^2 / M_{\rm Pl}$, where $g_*(T)$ denotes the effective number of relativistic degrees of freedom. Throughout this work, we will neglect the difference between the effective number of relativistic degrees of freedom contributing to the energy and entropy densities. Thermalization is dominated in the infrared since the portal operator is marginal and equilibrium is reached if $\Gamma_P / \Hubble$ is of order unity at the weak scale. This imposes the condition $|\lambda_P| \gtrsim 10^{-7}$. For a detailed analysis of thermalization through this portal, see, e.g., Ref~\cite{Bernal:2018kcw}. 

If thermalization is achieved, thermal effects stabilize the scalar at $S = 0$ at high temperatures since our focus is on the parameter space region where $\lambda_P > -3\lambda_S/2$ (see Eqs.~\eqref{eq:muT} and \eqref{eq:thermalcorrections}). This happens when the temperature is larger than the critical value 
\be
T_{\rm cr} = \sqrt{6} \sqrt{\frac{\lambda_P v_H^2 + 2 \lambda_S v_S^2}{2\lambda_P + 3\lambda_S}} \,.
\ee 
The quantities $v_H$ and $v_S$ appearing in the equation above are the zero temperature vevs for the Higgs doublet $H$ and the scalar singlet $S$, respectively, and their explicit expressions can be found in Eq.~\eqref{eq:vevs}. As the temperature drops below $T_{\rm cr}$, the field $S$ develops a non-vanishing vev. For this scenario, an important condition to be satisfied is having a reheating temperature after inflation $T_{\rm reh}$ larger than $T_{\rm cr}$. We express this condition in the limit $\lambda_P \ll \lambda_S$ and $v_H \ll v_S$ where we approximate $T_{\rm cr} \approx \sqrt{2} m_s/\sqrt{\lambda_S}$. Upon assuming instantaneous thermalization, the condition $T_{\rm reh} > T_{\rm cr}$ can be written as a lower bound on the Hubble parameter during inflation 
\be
    \Hubble_{\rm inf} \gtrsim 2.9 \times 10^{-8}\,{\rm GeV}\, \lambda_S^{-1} 
   \left[\frac{g_*(T_{\rm reh})}{107.75} \right]^{1/2}
    \left[\frac{m_s}{100\,{\rm TeV}} \right]^2 \,.
\ee
Here, we use as a reference value $g_*(T_{\rm reh}) = 107.75$ that accounts for the full SM bath contribution as well as the presence of $S$ in thermal equilibrium with the rest.

\subsection{DW Production II: Non-Thermal}

A drastically different scenario has the field $S$ never in thermal equilibrium neither with the SM bath nor with itself. This is the second scenario we study, and it requires that the couplings to SM fields and the abundance of $S$ or its self-couplings are sufficiently small. The formation of the DW network in this case depends on whether the symmetry was ever effectively restored in the early Universe, similar to the discussion of pre- and post-inflationary QCD axion scenarios. DWs are not formed if $v_S\gg \Hubble_{\rm inf}$ since our Hubble patch only experiences one vacuum. Contrarily to the QCD axion case, there is a non-vanishing potential for the field $S$ that can drive the field itself near the minimum if inflation lasts long enough. However, even if the vev of $S$ is much larger than the Hubble scale of inflation, some energy can still be stored in the field $S$ if $\Hubble_{\rm inf} \gg m_S$. 

In the opposite regime, $\Hubble_{\rm inf}\gg v_S$, the inflationary fluctuations assist transitions between the minima~\cite{Starobinsky:1994bd}. These transitions can be effective down to the smallest scales that exit the horizon during the primordial inflation. This restores the symmetry statistically and leads to the formation of DWs after inflation~\cite{Lyth:1992tw}. For this scenario, in addition to the requirement that $\Hubble_{\rm inf}\gg v_S$, the quartic coupling needs to be sufficiently large: Consider a patch of size $k_0^{-1}$ where $S = \pm v_S$. The probability that a patch of size $k^{-1}$ inside that patch transitions to the minimum at $S = \mp v_S$ is 
\be
    P_{\pm v_S \to \mp v_S} = 1-e^{-\Lambda (k/k_0)^3} \,, 
\ee
where~\cite{Maeso:2021xvl}
\be
    \Lambda \simeq \frac{\sqrt{\lambda_S}}{\sqrt{3}\pi^2} \exp\left[-\frac{2\pi^2 \lambda_S v_S^4}{3\Hubble_{\rm inf}^4}\right] \ .
\ee
For the transition probability to be higher than $50\%$ we need $\lambda_S \gtrsim 100 (k_0/k)^6$. This implies that, for example, patches that exited the horizon more than 2 $e$-folds before the end of inflation are populated fully randomly between the two minima if $\lambda_S \gtrsim 10^{-3}$.

Alternatively to the fluctuations, there can be a coupling between the singlet and the inflaton or the Ricci scalar that stabilizes $S=0$ during inflation and the DWs form at the end of the inflation when $S=0$ becomes a local maximum.

\subsection{DW Annihilation}

Regardless of the formation mechanism, the DW dynamics at early stages is determined by the surface tension of the walls defined as
\be
\sigma \equiv \int \td S \sqrt{2V} \ .
\ee
Here, the integral is computed between the two vacua. The DWs quickly reach the scaling solution~\cite{Press:1989yh,Garagounis:2002kt,Oliveira:2004he,Kawasaki:2014sqa} in which their energy density scales as $\rho_{\rm DW} \propto a^{-2}$ and the DW curvature radius is $\simeq 1/\Hubble(T)$. 

A small $\mathbb{Z}_2$-breaking bias term makes the DWs quasi-stable. The scaling solution holds as long as the energy gap $\Delta V$ between the two vacua is subdominant with respect to the tension $\sigma$.  Eventually, $\Delta V$ starts to drive the collapse of patches with a higher potential energy, and the DWs annihilate at the temperature $T_{\rm ann}$ that satisfies the following relation
\be \label{eq:Tann}
    \sigma \Hubble(T_{\rm ann}) \simeq \Delta V \,.
\ee

Evaluating $T_{\rm ann}$ requires the knowledge of both $\Delta V$ and $\sigma$. In the \emph{non-thermal} scenario the singlet is practically disconnected from the Higgs field and we can perform the DW computations purely in terms of the singlet parameters $\mu_S^2$, $\mu_3$ and $\lambda_S$. The \emph{thermal} case can be more complicated and, in general, the computations should be performed accounting for the dynamics of both $H$ and $S$. However, as we will see shortly, in the thermal case we are interested only in the regime where the mass of $s$ is much larger than the weak scale. Then, the formation of DWs happens at $T_{\rm cr} \gg v_H$ and, even if the electroweak phase transition happens before the DWs annihilate, the effect of the Higgs vev on the DW dynamics is negligible because $v_S \gg v_H$. Moreover, if the bias term is sufficiently small, the annihilation happens at $T_{\rm ann}\ll v_S$ and even the thermal corrections can be neglected. So, in both \emph{thermal} and \emph{non-thermal} cases the DW tension and the potential energy bias are given by
\bea \label{eq:sigmareal}
    \Delta V = & \, \frac23 \mu_3 v_S^3 \ , \\    
    \sigma = & \, \frac{2 \sqrt{2}}{3} \sqrt{\lambda_S} \, v_S^3 \ .
\eea
Thus the annihilation temperature results in
\be \label{eq:Tannreal}
    T_{\rm ann} \approx 23 \,{\rm TeV} \left[\frac{g_*(T_{\rm ann})}{100}\right]^{-1/4} \lambda_S^{-1/4} \left[\frac{\mu_3}{\rm eV}\right]^{1/2} \,.
\ee

Reaching the scaling solution before DWs annihilate is not guaranteed. Recent simulations~\cite{Chang:2023rll} indicate that this is achieved if $4\Delta V/(m_s \sigma) \lesssim 0.005$, and this happens for $\mu_3 \lesssim 0.002 \sqrt{\lambda_S} m_s$. In addition, in the \emph{thermal} case, we need to make sure that $\sigma \Hubble(T_{\rm cr}) \gg \Delta V$ as otherwise the DWs never form. As already mentioned in the paragraph above, we are interested in thermal DW formation only for $m_s \gg v_H$, and in this regime the constraint translates into an upper bound on the bias term
\be
    \mu_3 \ll 120\,{\rm eV} \frac{\sqrt{\lambda_S}}{2\lambda_P +3\lambda_S} \left[ \frac{g_*(T_{\rm cr})}{100} \right]^{1/2} \left[\frac{m_s}{100 \,{\rm TeV}} \right]^2 \ .
\ee
The above condition for $\mu_3$ guarantees that $T_{\rm ann}\ll v_S$.

If the annihilation is delayed for a long enough time, the DWs dominate over the radiation energy density, $\sigma \Hubble(T) > \rho_r(T)$. As discussed in Sec.~\ref{sec:GWs}, this is constrained by GW formation in the annihilation process.

\section{Gravitational waves}
\label{sec:GWs}

\begin{figure*}
\centering
\includegraphics[height=0.38\textwidth]{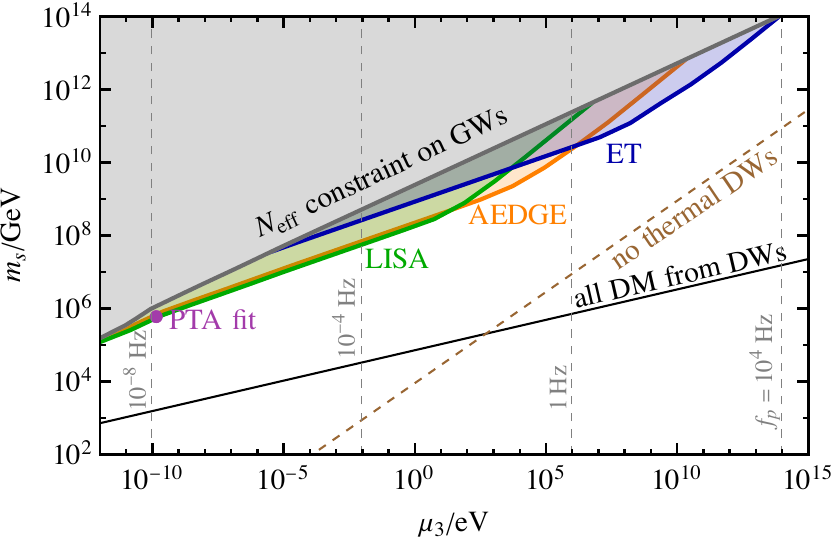}
\hspace{4mm}
\includegraphics[height=0.38\textwidth]{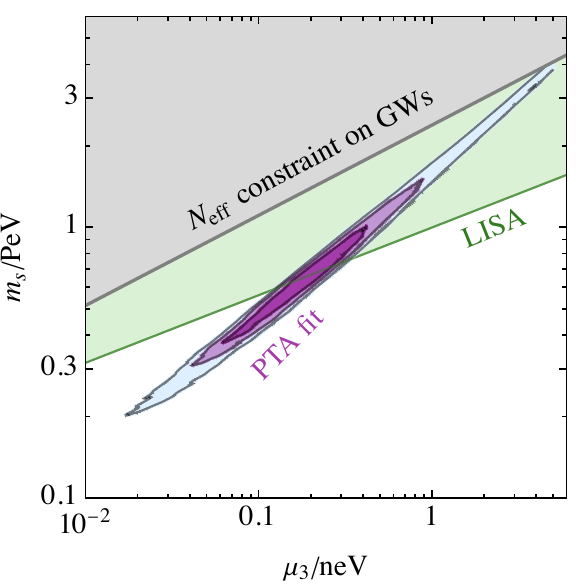}
\caption{\emph{Left panel:} The projected sensitivities (${\rm SNR}>8$) of LISA, AEDGE and ET on GW background generated by DWs in the real scalar singlet model with $\lambda_S = 0.1$. The vertical dashed contours indicate the peak frequency of the GW signal, $f_p\propto T_{\rm ann}$. Along the solid black line, the observed DM abundance in $s$ is generated from DW annihilation and the region above it is excluded by DM overproduction if $s$ is quasi-stable. The gray region is excluded by constraints from CMB and BBN on the total abundance of GWs and in the \emph{thermal} DW scenario the DWs form only in the region above the dashed brown line. \emph{Right panel:} A zoom of the left panel in the region of the PTA fit.}
\label{fig:GWs}
\end{figure*}

The long-lived DW network emits GWs (see~\cite{Saikawa:2017hiv} for a review). The resulting GW spectrum is
\be
    \Omega_{\rm GW}(f) = \Omega_{\rm GW,p} \!\left[ \frac14 \frac{\Omega_{\rm CT}(f_p)}{\Omega_{\rm CT}(f)} + \frac34 \frac{f}{f_p}\right]^{-1},
\ee
where $\Omega_{\rm GW,p}$ denotes the peak amplitude, $f_p$ the peak frequency, and the function $\Omega_{\rm CT}(f)$ is defined at the end of this paragraph. The current value of the peak amplitude $\Omega_{\rm GW,p}$ can be derived by starting from its value at the time when the DWs annihilate~\cite{Hiramatsu:2013qaa}
\be
	\Omega_{\rm GW,p}(T_{\rm ann}) = \frac{\epsilon_{\rm GW} \mathcal{A}^2 \sigma^2}{8 \pi M_{\rm Pl}^2 \rho_r(T_{\rm ann})} \,,
\ee
where $\rho_r(T)$ denotes the radiation energy density. The efficiency parameter is set to $\epsilon_{\rm GW}  \simeq  0.7$, and the area parameter to $\mathcal{A} \simeq 0.7$ as obtained from numerical simulations~\cite{Hiramatsu:2013qaa,Kawasaki:2014sqa}. Assuming instantaneous reheating after the DW annnihilation and using the condition in Eq.~\eqref{eq:Tann}, we can red-shift the above quantity and identify the peak amplitude today
\bea
   &\Omega_{\rm GW,p} = \Omega_{\rm GW,p}(T_{\rm ann}) \, \Omega_{r}(T_0) \, 
    \left[ \frac{g_*(T_0)}{g_*(T_{\rm ann})} \right]^{1/3}  \\ & \simeq 
    2.5 \times 10^{-9} \left[ \frac{100}{g_*(T_{\rm ann})} \right]^{1/3} \left[\frac{\sigma}{{\rm PeV}^3}\right]^4 \left[ \frac{{\rm GeV}^4}{\Delta V} \right]^{2} .
\label{eq:OmegaGWpeak}
\eea
Consistently with our conventions, we approximate the number of entropic degrees of freedom $g_{*s}(T)$ with the the number of effective degrees of freedom $g_*(T)$ contributing to the radiation energy density. For numerical values at the current temperature $T_0$, we set $g_*(T_0) \simeq 3.91$ and $\Omega_r(T_0) \simeq 5.4 \times 10^{-5}$.
The peak frequency is given by the horizon size at the time of the DW annihilation, corresponding to the present frequency of 
\bea
	f_p = & \, \frac{\Hubble(T_{\rm ann})}{2\pi} \left[\frac{a(T_{\rm ann})}{a(T_0)} \right] \\ &
 \simeq 26\, {\rm nHz}\, \left[ \frac{g_*(T_{\rm ann})}{100} \right]^{1/6} 
 \left[\frac{T_{\rm ann}}{\rm GeV}\right] \,.
\eea
Here, we introduce the Robertson–Walker scale factor $a(T)$ which is a function of the thermal bath temperature. For scales that were outside the horizon at the time of GW production, which correspond to frequencies $f<f_p$, the spectrum follows the causality tail~\cite{Franciolini:2023wjm,Ellis:2023oxs} scaling roughly as $\Omega_{\rm CT}(f) \propto f^3$~\cite{Caprini:2009fx,Domenech:2020kqm,Franciolini:2023wjm,Ellis:2023oxs}. At large frequencies, $f\gg f_p$, the spectrum falls as $\Omega_{\rm GW}(f) \propto f^{-1}$. 

For the real scalar singlet studied in this paper, using $v_s = m_s/\sqrt{2\lambda_S}$, we find the following parameters for the spectrum
\bea \label{eq:GWspectmodel}
    &\Omega_{\rm GW,p} \approx \frac{5.6 \times 10^{-34}}{\lambda_S} \!\left[ \frac{100}{g_*(T_{\rm ann})} \right]^{1/3} \!\left[ \frac{\rm eV}{\mu_3} \right]^{2} \!\left[ \frac{m_s}{100\,{\rm TeV}} \right]^6 , \\
    &f_p \approx 0.6 \,{\rm mHz} \left[ \frac{100}{g_*(T_{\rm ann})} \right]^{1/12} \lambda_S^{-1/4} \left[\frac{\mu_3}{\rm eV} \right]^{1/2} \,.
\eea

We estimate the parameter space regions that can be probed with the future GW experiments by computing the signal-to-noise ratio,
\be
    {\rm SNR} = \sqrt{\mathcal{T} \int \td f \left[\frac{\Omega_{\rm GW}(f)}{\Omega_n(f)}\right]^2} \,,
\ee
where $\mathcal{T}$ denotes the observation time and $\Omega_n(f)$ characterizes noise. In addition to the instumental noise of the detector, we include in $\Omega_n(f)$ the noise from galactic and extragalactic compact binary foregrounds as in~\cite{Lewicki:2021kmu}. In Fig.~\ref{fig:GWs} we show for $\lambda_S=0.1$ the regions where ${\rm SNR}>8$ in $\mathcal{T} = 4\,$yr is obtained with ET~\cite{Hild:2010id}, AEDGE~\cite{AEDGE:2019nxb,Badurina:2021rgt} and LISA~\cite{LISA:2017pwj,Robson:2018ifk}. The gray regions are excluded by the CMB and BBN constraints on the contribution of the total abundance of GWs, $\Omega_{\rm GW,tot} \approx 2\Omega_{\rm GW,p}$ on the effective number of relativistic degrees of freedom~\cite{Pagano:2015hma,Kohri:2018awv}. Notice that the dependence of the GW spectrum on $\lambda_S$ is much weaker than on $m_s$ and $\mu_3$. For example, reducing $\lambda_S$ by six orders of magnitude moves the boundary of the gray region to lower $m_s$ values by one order of magnitude and the peak frequency and, consequently, the sensitivity ranges of the GW experiments to higher $\mu_3$ values by three orders of magnitude.

The best fit DW interpretation of the PTA signal is obtained for $T_{\rm ann} \approx 0.85\,{\rm GeV}$ and $\alpha_* \equiv \sigma H(T_{\rm ann})/\rho_r(T_{\rm ann}) \approx 0.11$~\cite{Ellis:2023oxs} which translates to $\sigma^{1/3} \approx 0.91\,{\rm PeV}$. In the real singlet scalar model this corresponds to $\mu_3 \approx 0.4 \sqrt{\lambda_S} \,{\rm neV}$ and $v \approx 0.9 \lambda_S^{-1/6} \,{\rm PeV}$, giving $m_s \approx \lambda_S^{1/3} \,{\rm PeV}$. This point is indicated by the purple dot in the left panel of Fig.~\ref{fig:GWs} and in the right panel we show the 1, 2 and $3\sigma$ confidence level regions of the fit. The fit shows a high correlation between the model parameters because the fit is obtained mostly from the low-frequency tail of the spectrum where the signal remains roughly intact for constant $\Omega_{\rm GW}/f_0^3$. Interestingly, a large part of the parameter space that gives a good fit produces a GW background detectable with LISA. In the \emph{thermal} case, the formation of DWs that could produce the PTA signal requires that the reheating temperature after inflation is $T \gtrsim {\rm PeV} \, \lambda_S^{-1/6}$, whereas in the \emph{non-thermal} case the Hubble scale during inflation needs to be $\Hubble_{\rm inf}\gg m_s$ which, assuming instantaneous thermalization, corresponds to temperature $T \gtrsim 10^{6} \,{\rm PeV} \, \lambda_S^{1/3}$.

Recently, Refs.~\cite{Kitajima:2023cek,Ferreira:2024eru} found that the GW spectrum from DWs is enhanced compared to the result of~\cite{Hiramatsu:2013qaa} by more than an order of magnitude because of the GW production during the DW annihilation. This also shifts the peak to lower frequencies by more than a half. However, precise fitting factors were not provided in~\cite{Kitajima:2023cek,Ferreira:2024eru} so we have used the results of~\cite{Hiramatsu:2013qaa}. These extra factors would simply slightly shift our GW results shown in Fig.~\ref{fig:GWs} dominantly towards smaller $\mu_3$.

The solid black line in the left panel of Fig.~\ref{fig:GWs} shows the parameter space region where DW annihilations produce a DM abundance that matches the observed one (this requires that $s$ is enough long-lived and it is discussed in the next section, see Eq.~\eqref{eq:OmegasDW}). If $s$ is indeed stable on cosmological timescales, the amount of DM is overproduced above that line. Thus the parameter regime where the GW signal in the real singlet scalar model is detectable with PTAs, LISA, AEDGE or ET does not include a DM candidate but it is possible to extend the model e.g. with a fermion that interacts with $s$ and can play the role of DM~\cite{Zhang:2023nrs}. Moreover, interestingly, the DW annihilation can lead to formation of a large primordial black hole (PBH) abundance~\cite{Ferrer:2018uiu,Gelmini:2023ngs,Gouttenoire:2023ftk,Gouttenoire:2023gbn,Ferreira:2024eru,Dunsky:2024zdo}. In particular, the PTA result points to the region where a large abundance of $0.1 - 100 \,M_\odot$ PBHs might be generated in the DW network collapse~\cite{Gouttenoire:2023ftk,Gouttenoire:2023gbn,Ferreira:2024eru}. They can not constitute all DM~\cite{Carr:2020gox} but they provide a testable signatures. In particular, they form efficiently binaries whose GW signals can be searched with current and future GW detectors~\cite{Pujolas:2021yaw}. Instead, the asteroidal mass window, where PBHs can constitute all DM, corresponds to GW background from DW annihilation peaks at frequencies $0.01 \lesssim f_p/{\rm Hz} \lesssim 1$ and is in the range of LISA, AEDGE and ET~\cite{Ferreira:2024eru}.

\section{Dark matter}
\label{sec:DM}

We now turn to another important consequence of the formation of quasi-stable DW in the early Universe. The annihilation of the long-lived DW network produces $s$ particles that are dominantly non-relativistic. If the field $s$ is enough long-lived to survive until today, this is an additional DM production mechanism besides freeze-in and/or freeze-out mediated by the portal operator. 

In this section, we study the phenomenology of DM produced via DW annihilations. Using the exact scaling solution, the energy density of $s$ particles after the DW annihilation matches the one of the DWs just before they annihilate
\be
	\rho_s(T_{\rm ann}) \approx \rho_{\rm DW}(T_{\rm ann}) = \mathcal{A} \, \sigma \Hubble(T_{\rm ann}) \,,
\ee
where $\mathcal{A}$ denotes the area parameter. Then, the current energy density of $s$ can be obtained by rescaling its value at the DW annihilation epoch. We define $T_{\rm NR}$ as the temperature when $s$ particles enter the non-relativistic regime, and we introduce the parameter $\epsilon_s$ describing how energetic $s$ particles are after production via the relation $E_s \approx \epsilon_s m_s$. Thus, $s$ particles enter the non-relativistic regime when the value of the cosmic scale factor is $a(T_{\rm NR}) \simeq \epsilon_s a(T_{\rm ann})$. The rescaling of the $s$ energy density is achieved as follows
\bea \label{eq:rhos0}
    \rho_s(T_0) & \, = \rho_s(T_{\rm ann}) \left[ \frac{a(T_{\rm ann})}{a(T_{\rm NR})} \right]^4 
    \left[ \frac{a(T_{\rm NR})}{a(T_0)} \right]^3 \\ & \simeq 
    \,\frac{\mathcal{A}}{\epsilon_s} \left[ \frac{a(T_{\rm ann})}{a(T_0)} \right]^3 \sigma \Hubble(T_{\rm ann}) \ .
\eea
We use the value found in Ref.~\cite{Kawasaki:2014sqa} for $\epsilon_s$ and $\mathcal{A}$.\footnote{Recently, Ref.~\cite{Chang:2023rll} found a slightly different result for the DM abundance generated by the DWs of a complex scalar field with $\mathbb{Z}_2$ symmetry.} These were obtained from simulations of a complex scalar model with an approximate $\mathbb{Z}_N$ symmetry and suggest the value $\epsilon_s \approx 2$ independently of $N$. We assume that the result for $N=2$ approximately holds for the real scalar case, and this provides the value $\mathcal{A}\approx 0.7$. So, using Eq.~\eqref{eq:Tann} to relate $T_{\rm ann}$ to $\sigma$ and $\Delta V$, we get
\bea \label{eq:DMDW}
\Omega_s \approx & \, 7.1 \times 10^{-3} \left[\frac{100}{g_*(T_{\rm ann})}\right]^{1/4} \left[\frac{\sigma}{{\rm GeV}^3}\right]^{3/2} \left[\frac{{\rm eV}^4}{\Delta V}\right]^{1/2} \\
    \simeq & \, 0.1 \, \lambda_S^{-3/4} \left[\frac{100}{g_*(T_{\rm ann})}\right]^{1/4} \bigg[ \frac{m_s}{100\,{\rm TeV}} \bigg]^{3} \bigg[ \frac{\rm eV}{\mu_3} \bigg]^{1/2} ,
\eea
where in the second equality we plug the explicit expressions for $\sigma$ and $\Delta V$ in our model as given by Eq.~\eqref{eq:sigmareal}, and we use the relation $m_s = v_s/\sqrt{2\lambda_S}$. The condition needed to reproduce the observed DM abundance reads
\be
    \mu_3\simeq 0.2\,{\rm eV}\, \lambda_S^{-3/2} \, \left[\frac{100}{g_*(T_{\rm ann})}\right]^{1/2} \left[\frac{m_s}{100\,{\rm TeV}}\right]^6 \,.
\label{eq:OmegasDW}
\ee

The above constraint is identified by the solid black line in the left panel of Fig.~\ref{fig:GWs} (where we fix $\lambda_S = 0.1$ and evaluate $T_{\rm ann}$ self-consistently). This is a necessary condition to have DM accounted for by $s$ particles produced via DW annihilations, but it is not sufficient since we still do not know its lifetime. The relic density calculation to obtain $\Omega_s$ is independent of the portal coupling $\lambda_P$ that determines the mixing angle $\theta$ and, ultimately, the lifetime of $s$. If we require that these $s$ particles account for DM then we need a mixing angle $\theta$ small enough. This can happen only below the solid black line on the left panel of Fig.~\ref{fig:GWs}. DM would be overabundant above that line, and we would need to require a sufficiently large mixing angle to make sure that the overproduced $s$ particle decay early enough. 

Having a mixing angle such that $\tau_s$ equals the age of the Universe $t_0$ is in general not enough. Late-time decays of $s$ particles to SM final states, that are possible via the mixing with the Higgs doublet, inject energy during the cosmic expansion with an impact on the CMB anisotropy spectrum~\cite{Slatyer:2016qyl,Poulin:2016anj,Cang:2020exa} and produce particles that are detectable by ground- and space-based telescopes today (X-ray observations by XMM-Newton~\cite{Foster:2021ngm,Boyarsky:2006fg,Boyarsky:2006ag,Boyarsky:2007ay}, NuSTAR~\cite{Ng:2019gch,Roach:2019ctw,Roach:2022lgo} and INTEGRAL~\cite{Calore:2022pks}, and $\gamma$-ray and radio observations compiled in~\cite{Cirelli:2024ssz}). We discuss this limit in the final Sec.~\ref{sec:conclusions}, and we identify  allowed mixing angles with the green region in Fig.~\ref{fig:summary}. Here, we discuss mass bounds in the scenario where $s$ is stable enough to be a viable DM candidate and $\Omega_s \simeq 0.2$. They belong to this section because they are intrinsically connected to the DM production mechanism.

It is useful to rewrite the relic density constraint in Eq.~\eqref{eq:OmegasDW} in terms of the annihilation temperature 
\be \label{eq:msTann}
    m_s \simeq 4.6 \, {\rm TeV} \,\lambda_S^{1/3} \left[\frac{g_*(T_{\rm ann})}{100}\right]^{1/6} \left[\frac{T_{\rm ann}}{\rm GeV}\right]^{1/3} \,.
\ee
The DM abundance needs to be produced before matter-radiation equality, $T_{\rm ann} > T_{\rm eq} \simeq 3\,$eV, and this in turns implies the lower mass bound $m_s \gtrsim 4 \,{\rm GeV} \lambda_S^{1/3}$.

Imposing that DWs annihilate right before matter-radiation equality has another important consequence for the DM mass. Using our results, this corresponds to a GW background peak frequency $f_p \simeq 5\times 10^{-8}\,$nHz and an amplitude $\Omega_{\rm GW,p} \simeq 7\times 10^{-7} [\Omega_s/0.2]^2$. In this regime, the constraint from the CMB anisotropies on the GW abundance becomes relevant~\cite{Namikawa:2019tax}. This constraint can be approximated as $\Omega_{\rm GW,p} < 4\times 10^{-14} [f_p/10^{-8}\,{\rm nHz}]$. If the DM abundance originates from DW annihilation, this implies the stronger bound $m_s \gtrsim 9\,{\rm GeV}\, \lambda_S^{1/3}$.

An even stronger lower bound arises from the observations of the matter power spectrum. To our knowledge, the power spectrum of DM generated from DW annihilation has not been studied in detail in literature, and this would require numerical simulations. In what follows, we provide two estimates of the lower mass bound arising from the matter power spectrum observations. 

Large fluctuations in the energy density of $S$ are generated at the time of the DW annihilation. The modes mostly affected are those comparable to the horizon at that epoch with co-moving wavenumber $k_* \simeq a(T_{\rm ann})H(T_{\rm ann})$. The current Lyman-$\alpha$ forest constraints on the matter power spectrum extend up to $k_{{\rm Ly}\alpha}\approx 3 h\,{\rm Mpc}^{-1}$ and are in agreement with the cold DM model~\cite{Chabanier:2019eai}. As the first estimate, we require that the annihilation happens before the scale $k_{{\rm Ly}\alpha}$ re-enters horizon, $k_* > k_{{\rm Ly}\alpha}$. This gives $T_{\rm ann} > 230\,$eV and, consequently, using Eq.~\eqref{eq:msTann} we get 
\be
    m_s \gtrsim 16\,{\rm GeV}\, \lambda_S^{1/3} \,.
\ee

For the second estimate, we draw parallels with the case of the post-inflationary axion and assume the dimensionless matter power spectrum $\Delta$ to be
\be
    \Delta \approx A_{\rm peak} \min\left[ \frac{k^3}{k_*^3},\frac{k_*^n}{k^n} \right]\,,
\ee
where $A_{\rm peak}$ is the amplitude at the peak and $n>1$ is the spectral index in the UV part of the spectrum. The $\propto k^3$ scaling in IR corresponds to white noise fluctuations, as in the case of the post-inflationary axion. Then, following the approach of~\cite{Murgia:2019duy}, where an N-body numerical simulation was fed with an initial (linear) matter power spectrum that shares the same features of our case, we recast a lower bound
\be
    k_* \gtrsim \frac{300 h}{\rm Mpc} \left[ \frac{A_{\rm peak}}{0.001}\right]^{1/3} \,.
\ee
This gives
\be
    T_{\rm ann} \gtrsim 15\,{\rm keV}\left[ \frac{10}{g_*(T_{\rm ann})}\right]^{1/6} \left[ \frac{A_{\rm peak}}{0.001} \right]^{1/3} .
\ee
and, consequently,
\be
    m_s \gtrsim 25\,{\rm GeV} \lambda_S^{1/3} \left[\frac{g_*(T_{\rm ann})}{10}\right]^{1/9}\left[\frac{A_{\rm peak}}{0.001}\right]^{1/9} .
\ee
Notice that the dependence on the peak amplitude, whose value is difficult to estimate, is very weak.

For comparison, let us briefly discuss the scalar DM formation by the inflationary fluctuations themselves. Similarly as in the case of DM formation by DW annihilation, the scalar can be completely decoupled from the SM sector. For example, in the case of a decoupled quadratic potential, $V = m_s^2 s^2/2$, with $\Hubble_{\rm inf} \gg m_s$ the abundance of $s$ particles generated through inflationary fluctuations is~\cite{Starobinsky:1994bd}
\be \label{eq:missalign}
    \Omega_{s,{\rm inf}}\simeq 0.2 \left[\frac{m_s}{80{\rm GeV}}\right]^{-3/2} \left[\frac{\Hubble_{\rm inf}}{{\rm PeV}}\right]^{4} \,.
\ee
This scenario is very strongly constrained by the CMB bounds on DM isocurvature~\cite{Planck:2018jri} so that only in a small window around $m_s = 200$\,PeV all DM can be in $s$ particles and the isocurvature bound is avoided~\cite{Tenkanen:2019aij}. In the same way, the axion DM models where the Peccei-Quinn symmetry is not restored after inflation are strongly constrained by the isocurvature bound (see e.g.~\cite{Marsh:2015xka}). Same happens also in the case of a double well potential with $v_S\gg \Hubble_{\rm inf} \gg m_s$ because then only one of the minima is populated but the field has large fluctuations around that minimum. Instead, in the model with a double-well potential where the $s$ abundance is generated from DW annihilation, $s$ does not constitute an isocurvature component because it inherits the fluctuations in the DW annihilation temperature.

The quasi-stable $s$ relics produced from the DW annihilation constitute a perfectly viable DM candidate that escapes all direct, indirect and collider searches. This DM scenario is, therefore, similar to the model by Ref.~\cite{Fairbairn:2018bsw} where DM is produced through a non-minimal curvature coupling and no non-gravitational couplings are needed. In the case of DM production from DWs, even the non-minimal curvature coupling is not needed.

\section{Unavoidable Freeze-In}
\label{sec:thermal}

The last two sections discussed the phenomenological implications of the formation of quasi-stable DWs. For scalar mass values $m_s$ much larger than the weak scale we have a detectable GW stocastical background. However, in such a high mass region, the scalar field $s$ cannot be long-lived otherwise DW annihilation would produce too much DM. On the contrary, for lighter values of $m_s$ and mixing angle to make $s$ cosmologically stable, the lack of a GW signal is compensated by a viable DM production mechanism via DW annihilation. 

This section is devoted to discussing the phenomenology in the intermediate region where the mass of $s$ is much lighter than the ${\rm PeV}$ scale and its lifetime is at most of the order of the age of the Universe. Thus we cannot have a viable DM candidate in such a parameter space range, and we do not have a GW signal either. However, as we will show below, there is an unavoidable background of thermally produced $s$ particles via the freeze-in mechanism that can leave observable imprints. 

The idea here is similar to the irreducible axion background studied by Ref.~\cite{Langhoff:2022bij} where the authors highlighted the cosmological consequences of a light axion-like-particle (ALP) coupled to photons and electrons. Since ALP interactions with photons are mediated by an irrevelant operator, which leads to a UV dominated production, Ref.~\cite{Langhoff:2022bij} put a conservative upper bound of the reheat temperature after inflation $T_{\rm reh} \lesssim {\rm few} \, {\rm MeV}$ to preserve the success of BBN. This is not needed in our case since the interactions of the scalar field $s$ are all renormalizable and therefore the production is always IR dominated. The only assumption we make is that the radiation dominated phase extended back in time at least when the thermal bath temperature was of the order of $m_s$. Once one accepts it, what we discuss is a truly unavoidable freeze-in produced background without any further assumptions on $T_{\rm reh}$. 

A complete analysis of freeze-in production via the portal coupling $\lambda_P$ can be found in Ref.~\cite{Bernal:2018kcw}. The dominant production channel at small DM mass, $m_s < m_h/2$, is via decays of the SM-like Higgs boson. For larger DM mass, $m_s > m_h/2$, production takes place via bath particles annihilations mediated by the portal operator. For the former case the produced $s$ abundance scales as $\Omega_{s,{\rm FI}} \propto \lambda_P^2 \times m_s$, whereas at larger mass we have the scaling $\Omega_{s,{\rm FI}} \propto \lambda_P^2$. If $s$ would be stable, the observed DM abundance would be produced through freeze-in for the portal values $|\lambda_P| \approx 10^{-11} (m_s/{\rm GeV})^{-1/2}$ for $m_s < m_h/2$ and $|\lambda_P| \approx 3\times 10^{-11}$ for $m_s > m_h/2$. 

This unavoidable freeze-in production of $s$ and its subsequent decays are constrained by 
the success of BBN and by CMB predictions as well as by current X-ray observations. These impose constraints in the part of the parameter space where the freeze-in density is non-negligible. We follow the methodology of Ref.~\cite{Langhoff:2022bij} and define $\mathcal{F}_{s,{\rm FI}}$ as the fraction of DM that $s$ particles produced via freeze-in would constitute if they were absolutely stable. 

The constraints from BBN extend to lifetimes $\tau_s \gtrsim 1$\,s. We impose the bounds derived in~\cite{Kawasaki:2017bqm,Kawasaki:2020qxm} arising from photodissociation of light elements by injections of energetic photons and electrons by $s$ decays. The constraints are given as a lifetime dependent upper bound on $\mathcal{F}_{s,{\rm FI}}$
\be
    \mathcal{F}_{s,{\rm FI}} < \frac{\mathcal{F}^{\rm max}_{{\rm BBN},\gamma\gamma/e^+ e^-}(\tau_{s\to \gamma\gamma/e^+ e^-})}{1-e^{-t_0/\tau_s}} \,.
    \label{eq:birrBBN}
\ee
where the denominator accounts for the fraction that has decayed by today. The binding energy of the deuterium implies that this constraint holds only for $m_s > 4.4\,{\rm MeV}$.

Moving forward across the expansion history, CMB spectral distortions~\cite{Kawasaki:2020qxm} (see also e.g.~\cite{Chluba:2013wsa,Dimastrogiovanni:2015wvk,Balazs:2022tjl}) put a lifetime dependent upper bound
\be
    \mathcal{F}_{s,{\rm FI}} < \frac{\mathcal{F}^{\rm max}_{\rm SD}(\tau_s)}{1-e^{-t_0/\tau_s}} \,.
    \label{eq:birrSD}
\ee
If we account for the $s\to \gamma\gamma$ and $s\to e^+ e^-$ channels, the anisotropy spectrum of CMB puts the bound
\be
    \mathcal{F}_{s,{\rm FI}} < \frac{\tau_{s\to \gamma\gamma/e^+ e^-}}{t^{\rm min}_{{\rm CMB},\gamma\gamma/e^+ e^-}} \exp\!\left(\frac{t_{\rm CMB}}{\tau_s}\right)^{2/3} \,,
    \label{eq:birrCMB}
\ee 
where $t^{\rm min}_{{\rm CMB},\gamma\gamma/e^+ e^-}$ arises from Planck observations of CMB anisotropies~\cite{Slatyer:2016qyl,Poulin:2016anj,Cang:2020exa}. 

Finally, X-ray satellites provides us with the bound
\be
\mathcal{F}_{s,{\rm FI}} < \frac{\tau_{s\to \gamma\gamma}}{t^{\rm min}_{\rm DM}} \exp\!\left(\frac{t_0}{\tau_s}\right)
\label{eq:birrXray}
\ee
where $t^{\rm min}_{\rm DM}$ arises from XMM-Newton~\cite{Foster:2021ngm,Boyarsky:2006fg,Boyarsky:2006ag,Boyarsky:2007ay}, NuSTAR~\cite{Ng:2019gch,Roach:2019ctw,Roach:2022lgo} and INTEGRAL~\cite{Calore:2022pks} observations. We will comment in the conclusions how the freeze-in region of this model is not subject to constraints from observations of gamma rays.

\section{Conclusions}
\label{sec:conclusions}

We have studied in this paper the most minimal extension of the SM where only a gauge singlet real scalar $S$ is added to the field content. In spite of such a minimal framework, the cosmological implications we have found are rather rich and diverse. We have analyzed the DW phenomenology and its implications for GW and DM, and we have identified an unavoidable freeze-in contribution that can leave a trace of its presence.

\begin{figure*}
\centering
\includegraphics[width=0.85\textwidth]{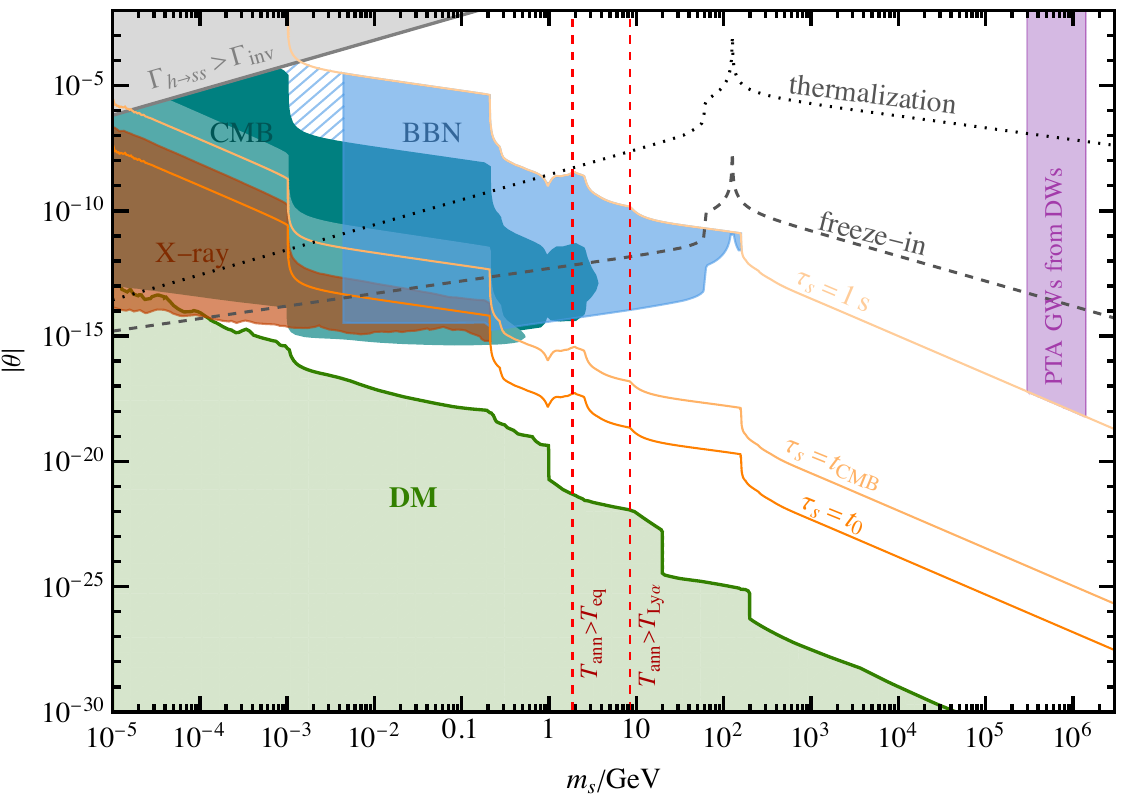}
\caption{Summary plot in the $(m_s, |\theta|$) plane at fixed $\lambda_S = 0.1$. The {\color{Orange} \textbf{orange}} curves show isocontours of constant lifetime $\tau_s$. Along the {\color{Black} \textbf{black}} dashed curve, the freeze-in abundance of $s$ would equal the observed DM abundance if $s$ was stable, and above the {\color{Black} \textbf{black}} dotted curve $s$ thermalizes with the SM bath. The {\color{Gray} \textbf{gray}} region is excluded by the invisible decays of the Higgs boson. In the {\color{RedViolet} \textbf{purple}} region, the $s$ mass is in the range where the DW annihilation can generate the GW background observed by the PTAs and $s$ decays before BBN. In the {\color{OliveGreen} \textbf{green}} region $s$ lifetime exceeds the constraints arising from X-ray, CMB, $\gamma$-ray and radio observations and $s$ can constitute DM. The {\color{Red} \textbf{red}} dashed vertical lines identify DM lower mass bounds: the DW annihilation happens before matter radiation equality $T_{\rm eq}$, and before the temperature $T_{{\rm Ly}\alpha}$ when the scales relevant for the Lyman-$\alpha$ forest constraints re-enter the horizon. Finally, the unavoidable freeze-in production puts the following bounds: the {\color{RoyalBlue}\textbf{blue}} region is excluded by the BBN constraints, the {\color{TealBlue} \textbf{turquoise}} regions are excluded by the constraints from CMB anisotropies (lighter) and spectral distortions (darker), the {\color{Mahogany} \textbf{brown}} region is excluded by X-ray constraints.}
\label{fig:summary}
\end{figure*}

We summarize our main findings in Fig.~\ref{fig:summary} where we plot the experimental and observational constraints in the $(m_s, |\theta|)$ plane. This figure neglects the $\mathbb{Z}_2$-breaking terms and investigates the phenomenology in terms of the three parameters appearing in the scalar potential in Eq.~\eqref{eq:VZ2}. We fix $\lambda_S = 0.1$, and the mass $m_s$ and the angle $\theta$ in the small mixing limit are expressed in terms of the potential parameters in Eqs.~\eqref{eq:spectrum} and \eqref{eq:angle}, respectively. 

The solid orange lines corresponds to constant lifetime, and we identify the curves where $\tau_s$ equals one second, the time of CMB formation $t_{\rm CMB}$, and the present age of the Universe $t_0$.

The dotted gray line is identified by the thermalization condition, and above such a line the scalar field $s$ reaches thermal equilibrium with the primordial bath. The dashed gray line has a similar shape, just slightly shifted below, and corresponds to the region where the comoving abundance of $s$ (if stable) matches the observed DM value. The peak in both curves is at $m_s \simeq m_h$ where the mixing angle is maximal (see Eq.~\eqref{eq:theta}). Notice that $\Omega_{s,{\rm FI}}\propto \theta^2$ so contours of constant $\Omega_{s,{\rm FI}}$ can be obtained trivially by shifting vertically the one already shown. We see that for $\lambda_S=0.1$ the freeze-in mechanism produces an amount of $s$ particles that matches the measured abundance today only for $m_s \lesssim 10$\,MeV. However, this region is excluded by the bounds listed in Sec.~\ref{sec:thermal} and discussed in the next paragraphs. Only for $m_s \lesssim 100$\,keV freeze-in production is a viable mechanism to account for all the observed DM abundance, and this corresponds to mixing angle values around $\theta \sim 10^{-15}$. Notice that the warm DM constraints impose the lower bound $m_s \gtrsim 10\,{\rm keV}$ on freeze-in produced DM~\cite{Garzilli:2019qki,DEramo:2020gpr,Villasenor:2022aiy,Chatterjee:2023mlh}.

In the \emph{thermal} DW formation scenario, the portal coupling is such that we are in a parameter space region above the dotted gray line. For $m_s > 1\,$GeV, this implies a lifetime that is much shorter than one second and therefore the scalar field decays before BBN. On the opposite mass range ($m_s < 1\,$GeV), if we keep $\lambda_P$ to be perturbative then the resulting freeze-out density of $s$ is larger than the observed DM abundance (see e.g.~\cite{Bernal:2018kcw}). Thus if we consider the \emph{thermal} DW formation scenario then $s$ can not constitute DM. The regime $\lambda_P\gtrsim 0.01$ and $m_s < m_h/2$ is, in addition, excluded by the invisible decay of the Higgs boson~\cite{Cline:2013gha}, and this constraint is shown in gray in Fig.~\ref{fig:summary}.

In the \emph{non-thermal} DW formation scenario, where the portal coupling is rather small ($|\lambda_P| < 10^{-7}$), scalars $s$ can be produced both via DW annihilations and the freeze-in mechanism~\cite{Hall:2009bx,Bernal:2017kxu}. However, as we already explained, we have a detectable GW signal only if the scalar field is short-lived and therefore cannot constitute the observed DM abundance. This is easily understood once one connects the spectral parameters of the GW background with the the abundance of $s$ particles generated by DW annihilations. The GW peak amplitude $\Omega_{\rm GW,p}$ and frequency $f_p$ are provided by Eq.~\eqref{eq:GWspectmodel} whereas the DM abundance $\Omega_s$ can be found in Eq.~\eqref{eq:DMDW}. Upon combining these results we find  
\be
    \Omega_{\rm GW,p} \approx 10^{-21}\, \left[\frac{\Omega_s}{0.2}\right]^2 \left[\frac{f_p}{\rm nHz}\right]^{-2} \ .
\ee
In the range of frequencies that can be probed with pulsar timing arrays and interferometric GW observatories, the GW energy density is detectable if $\Omega_s \gg 0.2$ and therefore the field $s$ has to decay quickly enough.

The parameter space region relevant to GW signals is on the right of Fig.~\ref{fig:summary}. A large GW signal can be generated in the real scalar singlet model if the lifetime of $s$ is small $\tau_s \lesssim 1\,$s, so that it decays before BBN. This requires a large enough portal coupling so the $s$ decays quickly but not too large to be consistent with Higgs bounds. As shown in Fig.~\ref{fig:summary}, the lower bound on the mixing angle is quite weak: for $m_{\rm s} \gtrsim 100\,$TeV the $s$ lifetime is $\tau_s\lesssim 1\,$s if $|\theta| \gtrsim 10^{-17}$. Consequently, the regime with large GW signal can be realized in both the \emph{thermal} and \emph{non-thermal} DW formation scenarios.

The GW background observed recently in the pulsar timing array data can be explained by the GWs generated by the domain walls if the singlet mass is $m_s\simeq \lambda_S^{1/3}\,$PeV with the cubic bias term $\mu_3 \simeq 0.4\sqrt{\lambda_S}\,$neV. The reheat temperature after inflation in this case needs to exceed $T_{\rm reh} \gtrsim {\rm PeV} \, \lambda_S^{-1/6}$ or $T_{\rm reh} \gtrsim 10^{6} \,{\rm PeV} \, \lambda_S^{1/3}$ depending whether the domain walls formed because of thermal corrections or fluctuations of the singlet during inflation.

DW annihilations produce the observed DM abundance if $\mu_3/{\rm eV}\simeq 0.2\lambda_S^{-3/2}[m_s/(100\,{\rm TeV})]^{6}$. There are further constraints that need to be satisfied. Red dashed vertical lines in Fig.~\ref{fig:summary} identifies lower mass bounds as dictated from Lyman-$\alpha$ forest observations. Furthermore, the singlet has to be long-lived enough to constitute DM and this imposed a mixing angle $|\theta| < 10^{-22}$ for $m_s \gtrsim 10\,$GeV. There are also late-time decays of $s$ that are constrained by galactic and extragalactic background observations, compiled in~\cite{Cadamuro:2011fd,Langhoff:2022bij}. In Fig.~\ref{fig:summary} we show in green the region allowed by these constraints. In this region $s$ can constitute all DM and it can be generated from the DW annihilation. In particular, $s$ has no detectable coupling to  SM particles and it does not have a non-minimal coupling to gravity. This DM candidate escapes all direct and indirect searches.

If DW annihilation leaves neither detectable GW nor stable DM particles, we can still put bounds on this framework via the unavoidable freeze-in production. The blue region in Fig.~\ref{fig:summary} is excluded by the BBN bound quantified by Eq.~\eqref{eq:birrBBN}. Likewise, the dark and light turquoise regions are excluded by CMB spectral distortions and spectral anisotropy bounds given by the conditions in Eqs.~\eqref{eq:birrSD} and \eqref{eq:birrCMB}, respectively. The former extends to $s$ lifetimes $\tau_s \simeq 3\times 10^5$\,s. The remaining region between the BBN and CMB constraints is not allowed either because in that regime $s$ freezes out while relativistic and decays to radiation after $1$\,s adding to the effective number of neutrino species $N_{\rm eff}$. This change by $\Delta N_{\rm eff} > 0.3$ is excluded by BBN~\cite{Fields:2019pfx} and CMB~\cite{Planck:2018vyg} observations. Finally, the brown region is excluded by the X-ray bounds given by Eq.~\eqref{eq:birrXray}. We notice how the freeze-in region is not subject to bounds from gamma rays because having $s$ in that mass range and enough long-lived to survive until today requires mixing angles that are too small to give a detectable signal.

All the bounds reported in our analysis are obtained by existing experiments and surveys. It would be important to explore how the upcoming experimental program - on several frontiers - is going to change the shape of our model, that despite its simplicity has such a vast and rich phenomenology. We believe that this could be an interesting update of our work.

\begin{acknowledgments}
We thank Oriol Pujol\`as and Edoardo Vitagliano for insightful discussions. This work is supported in part by the Italian MUR Departments of Excellence grant 2023-2027 ``Quantum Frontiers''. The work of F.D. is supported by Istituto Nazionale di Fisica Nucleare (INFN) through the Theoretical Astroparticle Physics (TAsP) project. F.D. acknowledges support from the European Union’s Horizon 2020 research and innovation programme under the Marie Skłodowska-Curie grant agreement No 860881-HIDDeN. The work of V.V. was supported by the European Union's Horizon Europe research and innovation programme under the Marie Sk\l{}odowska-Curie grant agreement No 101065736, and by the Estonian Research Council grants PRG803, RVTT3 and RVTT7 and the Center of Excellence program TK202.
\end{acknowledgments}

\bibliography{refs}
\end{document}